\documentclass{article}
\usepackage{ijcai21}

\usepackage{times}

\usepackage{soul}
\usepackage{url}
\usepackage[hidelinks]{hyperref}
\usepackage[utf8]{inputenc}
\usepackage[small]{caption}
\usepackage{enumitem}
\usepackage{graphicx}
\usepackage{amsmath}
\usepackage{booktabs}
\usepackage{color}
\usepackage{subfigure}
\usepackage{multirow}
\usepackage{makecell}
\newtheorem{myDef}{Definition}

\newcommand{\nosection}[1]{\smallskip\noindent\textbf{#1.}}

\title{Cross-Domain Recommendation: Challenges, Progress, and Prospects}

\author{
Feng Zhu$^1$\and
Yan Wang$^2$ \and
Chaochao Chen$^{1}$\footnote{Contact Author}\and
Jun Zhou$^1$\and
Longfei Li$^1$\and
Guanfeng Liu$^2$
\affiliations
$^1$ Ant Group, Hangzhou 310012, China\\
$^2$ Department of Computing, Macquarie University, Sydney, NSW 2109, Australia\\
\emails
\{zhufeng.zhu, chaochao.ccc, jun.zhoujun, longyao.llf\}@antgroup.com \\
\{yan.wang, guanfeng.liu\}@mq.edu.au
}

\begin{document}
	
	\maketitle
	
	\begin{abstract}
		
		To address the long-standing data sparsity problem in recommender systems (RSs), cross-domain recommendation (CDR) has been proposed to leverage the relatively richer information from a richer domain to improve the recommendation performance in a sparser domain. 
		Although CDR has been extensively studied in recent years, there is a lack of a systematic review of the existing CDR approaches. 
		To fill this gap, in this paper, we provide a comprehensive review of existing CDR approaches, including challenges, research progress, and prospects. 
		Specifically, we first summarize existing CDR approaches into four types, including single-target CDR, multi-domain recommendation, dual-target CDR, and multi-target CDR. 
		We then present the definitions and challenges of these CDR approaches. 
		Next, we propose a full-view categorization and new taxonomies on these approaches and report their research progress in detail. 
		In the end, we share several promising prospects in CDR.
	\end{abstract}
	
	\section{Introduction}\label{Intro}

In the past couple of decades, recommender systems (RSs) have become a popular technique in many web applications, e.g., \emph{Youtube} (video sharing), \emph{Amazon} (e-commerce), and \emph{Facebook} (social networking), as they provide suggestions of items to users so that they can avoid facing the information overload problem \cite{ricci2015recommender}. In the existing RSs, collaborative filtering (CF) has been proven to be the most promising technique \cite{chen2018semi}. The general idea of CF techniques is to recommend items to a user based on the observed preferences of other users whose historical preferences are similar to those of the target user.

\nosection{Motivation of Cross-Domain Recommendations}
In most real-world application scenarios, few users can provide ratings or reviews for many items \cite{ricci2015recommender} (i.e., data sparsity), which reduces the recommendation accuracy of CF-based models. Almost all existing CF-based recommender systems suffer from this long-standing data sparsity problem to some extent, especially for new items or users (the cold-start problem). This problem may lead to over-fitting when training a CF-based model, which significantly reduces recommendation accuracy. To address the data sparsity problem, \emph{cross-domain recommendation} (CDR) \cite{berkovsky2007cross} has emerged to utilise the relatively richer information, e.g., user/item information \cite{chung2007integrated}, 
thumbs-up \cite{shapira2013facebook}, tags \cite{fernandez2014exploiting}, reviews \cite{tan2014cross}, and observed ratings \cite{zhu2018deep}, from the richer (source) domain to improve the recommendation accuracy in the sparser (target) domain. For example, the Douban's\footnote{Douban URL: https://www.douban.com} RS can recommend books to users according to their movie reviews (i.e., CDR), since a common user in different domains is likely to have similar tastes.

\begin{figure*}[t]
 	\begin{center}
 	 	\subfigure[Content-level relevance]{
 			\label{fig:subfig:STCDR-CLR}
 			\includegraphics[width=0.32\textwidth]{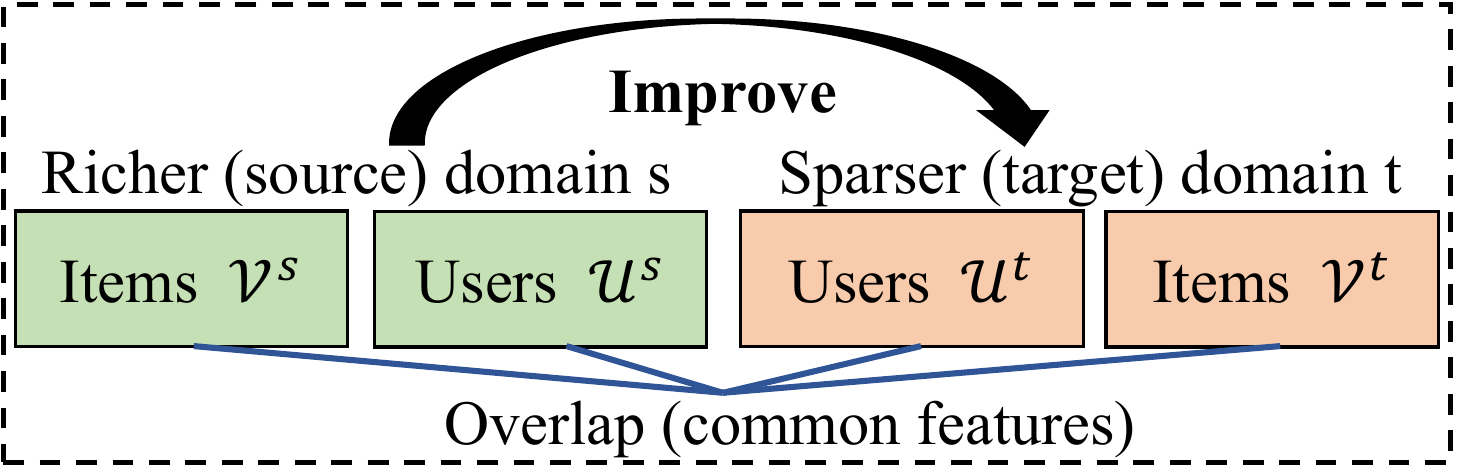}}
 		 \subfigure[User-level relevance]{
 			\label{fig:subfig:STCDR-ULR}
 			\includegraphics[width=0.32\textwidth]{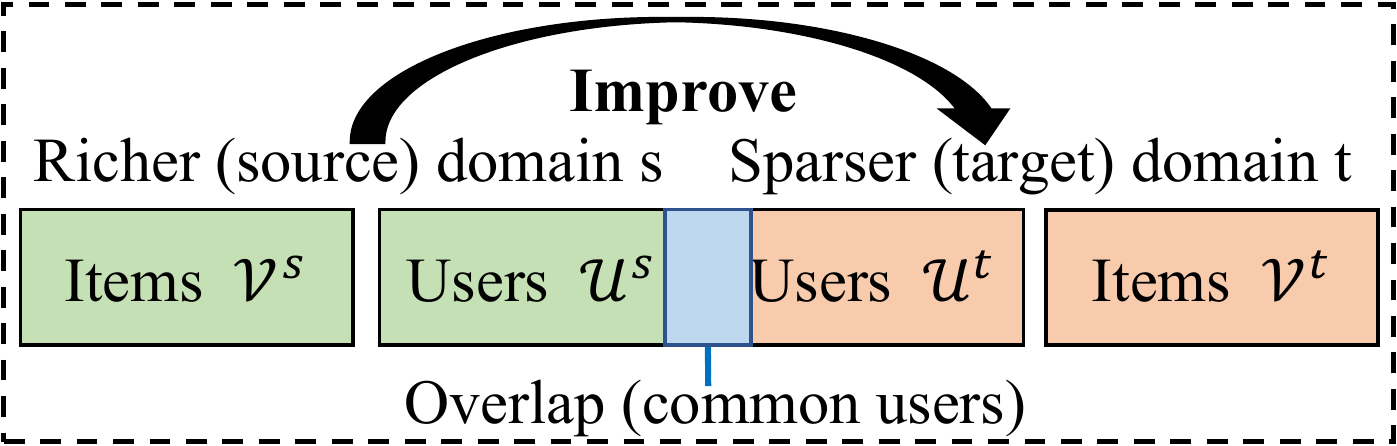}}
 		 \subfigure[Item-level relevance]{
 			\label{fig:subfig:STCDR-ILR}
 			\includegraphics[width=0.32\textwidth]{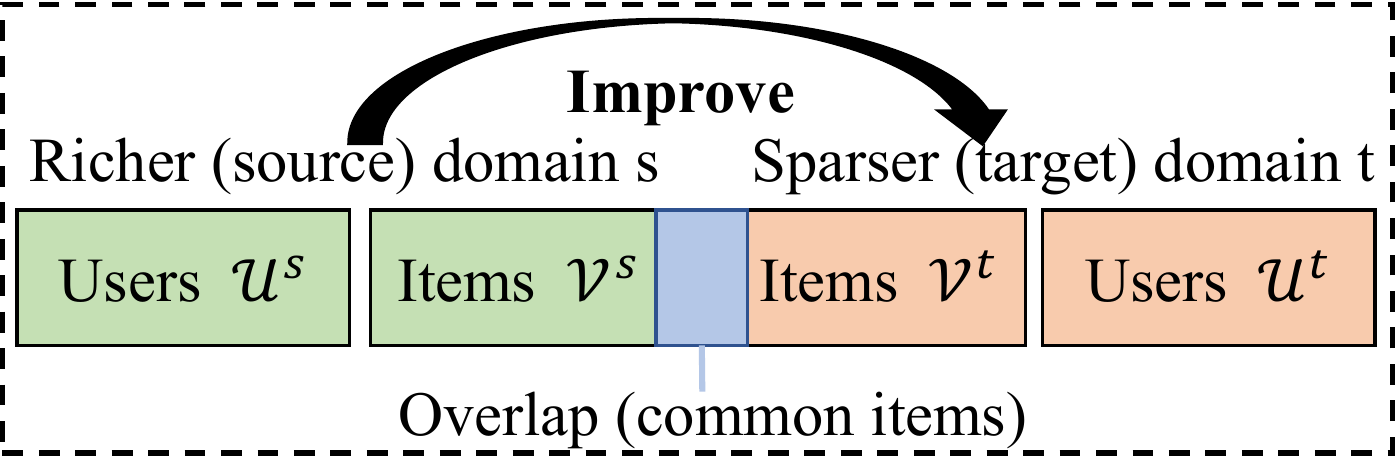}}
 	\caption{Single-target CDR scenarios}
 	\label{STCDR_Scenarios}
 	\end{center}
 \end{figure*}

\nosection{Notion of Domain} 
In the literature, researchers have attempted to define the notion of domain. However, the concept of domain is still very confusing. For example, some of them treated items (e.g., movies and books) as different domains \cite{hu2013personalized}, while others considered items in different sub-categories, e.g., such as textbooks and novels, as different domains \cite{cao2010transfer}. These definitions mainly focus on the difference of items but the difference of users is ignored. In this survey paper, we define domain from the following three perspectives, i.e., content-level relevance, user-level relevance, and item-level relevance.
\begin{itemize}[leftmargin=*]
    \item \textbf{Content-level relevance}. In the dual/multiple domains, there are the same content or metadata features, e.g., keywords and tags, from user preferences and item details. However, there are not common users/items in these domains, e.g., Amazon music\footnote{Amazon music URL: https://www.amazon.com/music} and Netflix\footnote{Netflix URL: https://www.netflix.com}.
    \item \textbf{User-level relevance}. In the dual/multiple domains, there are common users and different levels of items -- such as \textit{attribute-level} (i.e., items in the same type (e.g., book) with different attribute values, e.g., textbooks and novels) and \\textbf{type-level} (i.e., items in different types, e.g., movies and books).
    \item \textbf{Item-level relevance}. In the dual/multiple domains, there are common items (e.g., movies) and different users, e.g., the users in MovieLens\footnote{MovieLens URL: https://movielens.org} and Netflix systems. These users are totally different or it is difficult to distinguish the overlapped users among different recommender systems. In the literature \cite{zhu2018deep,zhu2020graphical}, this type of cross-domain recommendation is also referred to as `cross-system recommendation'.
\end{itemize}
\nosection{Classification of CDR}
Conventional CDR approaches can be generally classified into three groups: (1) content-based transfer approaches, (2) embedding-based transfer approaches, and (3) rating pattern-based transfer. \emph{Content-based transfer} mainly handles the CDR problems with content-level relevance and tends to link different domains by identifying similar content information, such as item details \cite{winoto2008if}, user-generated reviews \cite{tan2014cross}, and social tags \cite{fernandez2014exploiting}. In contrast, \emph{embedding-based transfer} \cite{zhu2020graphical} mainly handles the CDR problems with user-level relevance and item-level relevance. This class involves first training different CF-based models (such as 
singular value decomposition \cite{deerwester1990indexing}, maximum-margin matrix factorisation \cite{srebro2005maximum}, 
 probabilistic matrix factorisation \cite{mnih2008probabilistic}, 
 bayesian personalised ranking \cite{rendle2009bpr}, 
neural collaborative filtering \cite{he2017neural} and deep matrix factorisation \cite{xue2017deep}) 
and graphic models  \cite{zhu2020graphical,liu2020cross}) to obtain user/item embeddings, and then transferring these embeddings through common or similar users/items across domains. Different from embedding-based transfer, \emph{rating pattern-based transfer} tends to transfer an independent knowledge, i.e., rating pattern s, across domains. In contrast to the content-based transfer approaches, embedding-based and rating pattern-based transfer approaches typically employ machine learning techniques, such as multi-task learning \cite{singh2008relational}, transfer learning \cite{zhang2016multi}, clustering \cite{farseev2017cross}, and neural networks \cite{zhu2018deep}, to transfer knowledge across domains.

The above conventional CDR approaches are single-target approaches that can only leverage the auxiliary information from a richer domain to help a sparser domain. However, each of the domains may be relatively richer in certain types of information (e.g., ratings, reviews, user profiles, item details, and tags); if such information can be leveraged well, it is possible to improve the recommendation performance in all domains simultaneously rather than in a single target domain only. To this end, dual-target CDR \cite{zhu2019dtcdr,li2019ddtcdr,zhu2020graphical,liu2020cross} and multi-target CDR \cite{cui2020herograph} have been proposed recently to improve the recommendation performance in dual/multiple domains. 

\nosection{The Motivation of This Survey} 
CDR is not a new research area, and in the literature, there are two survey papers \cite{fernandez2012cross,cantador2014tutorial} and a systematical handbook \cite{ricci2015recommender} which have carefully introduced and analyzed this area. However, after these tutorials, there are many new challenges, e.g., feature mapping, embedding optimization, and negative transfer,  and new directions, e.g., dual-target CDR and multi-target CDR. These new research trends motivate us to analyze the challenges in CDR and summarize the research progress.

\nosection{Our Contributions} 
The main contributions of this survey paper are summarized as follows:
\begin{itemize}[leftmargin=*]
    \item We provide a detailed overview of the challenges in CDR and classify them from a developing perspective, which provides a whole view of the development in CDR area.
    \item We present a comprehensive review of current research progress in CDR. Specifically, we analyze the contributions of existing approaches and the similarities and differences between them.
    \item We outline some promising future research directions in CDR, which would shed light on the development of the research community.
\end{itemize}
	\section{Problems and Challenges}\label{Section_PC}
Cross-domain recommendation problem has been formulated in different recommendation scenarios, i.e., single-target CDR, multi-domain recommendation, dual-target CDR, and multi-target CDR. The main differences among these scenarios are the scales of domains, overlaps, and improvement targets. In this section, we introduce these particular CDR scenarios and their corresponding challenges.

\subsection{Single-Target CDR}
Single-target CDR is a conventional recommendation scenario in CDR area and most of the existing CDR approaches focus on this scenario. We define this recommendation problem as follows.
\begin{myDef} \label{Definition_STCDR}
 	\textbf{Single-Target Cross-Domain Recommendation}: Given the source domain $s$ (including a user set $\mathcal{U}^s$ and an item set $\mathcal{V}^s$) with richer data --- such as explicit feedback (e.g., ratings and comments), implicit feedback (e.g., purchase and browsing histories), and side information (e.g., user profiles and item details) --- and the target domain $t$ (including a user set $\mathcal{U}^t$ and an item set $\mathcal{V}^t$) with sparser data, single-target CDR is to improve the recommendation accuracy in $t$ by leveraging the auxiliary information in $s$.
\end{myDef}
 
As introduced in Section \ref{Intro}, we define the notion of domain from three perspectives, i.e., content-level relevance, user-level relevance, and item-level relevance.
Thus, single-target CDR scenario is divided into three corresponding sub-scenarios as well (see Figure \ref{STCDR_Scenarios}).
We summarize three main challenges for the single-target CDR scenarios. 

\nosection{Building content-based relations (CH1)} 
In Figure \ref{fig:subfig:STCDR-CLR}, to improve the recommendation accuracy in the target domain, one should first build content-based relations, then choose similar users/items according to their common features, and finally, transfer/share other features between similar users/items across domains. Therefore, how to build a suitable content-based relation in single-target CDR (content-level relevance) scenario is very important and challenging.

\nosection{Generating accurate user/item embeddings or rating patterns (CH2)} In Figures \ref{fig:subfig:STCDR-ULR} and \ref{fig:subfig:STCDR-ILR}, to improve the recommendation accuracy in the target domain, one should first generate accurate user/item embeddings or rating patterns, and then transfer/share the embeddings of common users/items or rating patterns of common users across domains. Therefore, how to generate accurate embeddings or rating patterns is a fundamental and crucial challenge.

\nosection{Learning accurate mapping relations (CH3)} 
For the three scenarios of single-target CDR in Figure \ref{STCDR_Scenarios}, a naive transfer strategy is to directly replace the features/embeddings of users/items in the target domain with those of their similar users/items in the source domain \cite{zhao2017unified}. 
This strategy is simple but not intelligent. 
An elegant way is to first learn accurate mapping relations between two domains, and then transfer the knowledge (e.g., user/item embeddings and rating patterns) learned from a source domain to a target domain according to the learned mapping relations. 
Following such an intuition, how to learn accurate mapping relations becomes a crucial challenge.
 \begin{figure}[t]
 	\begin{center}
 		\includegraphics[width=\columnwidth]{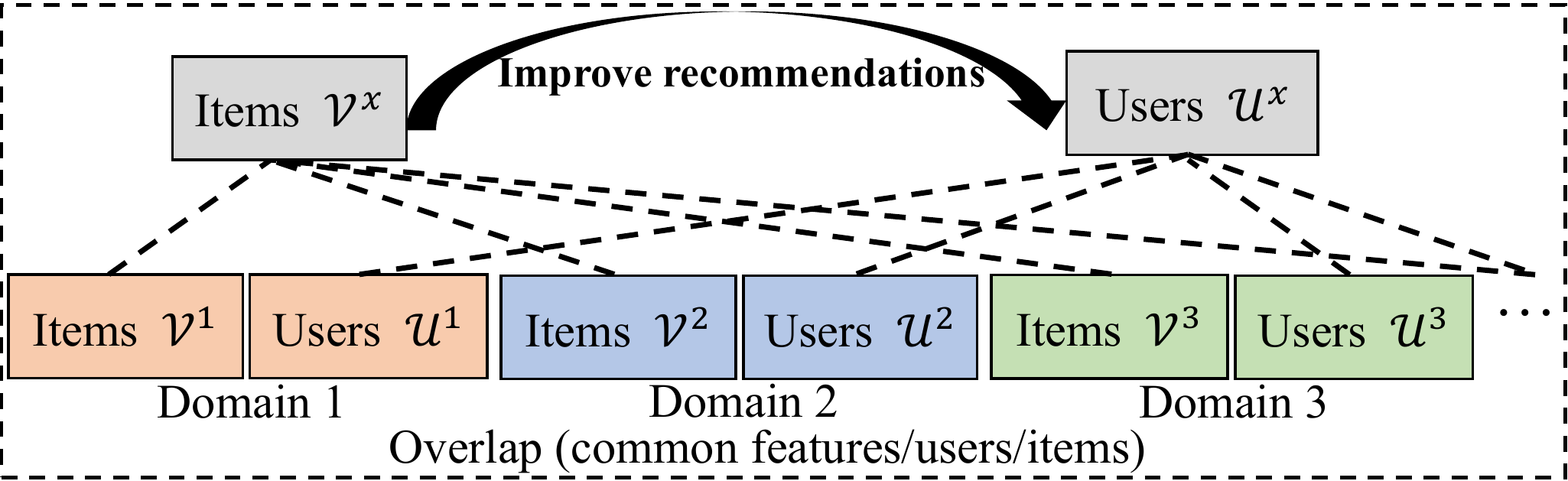}
 		\caption{MDR scenario}\label{MDR}
 \end{center}
\end{figure}
  \begin{figure}[t]
 	\begin{center}
 		\includegraphics[width=0.35\textwidth]{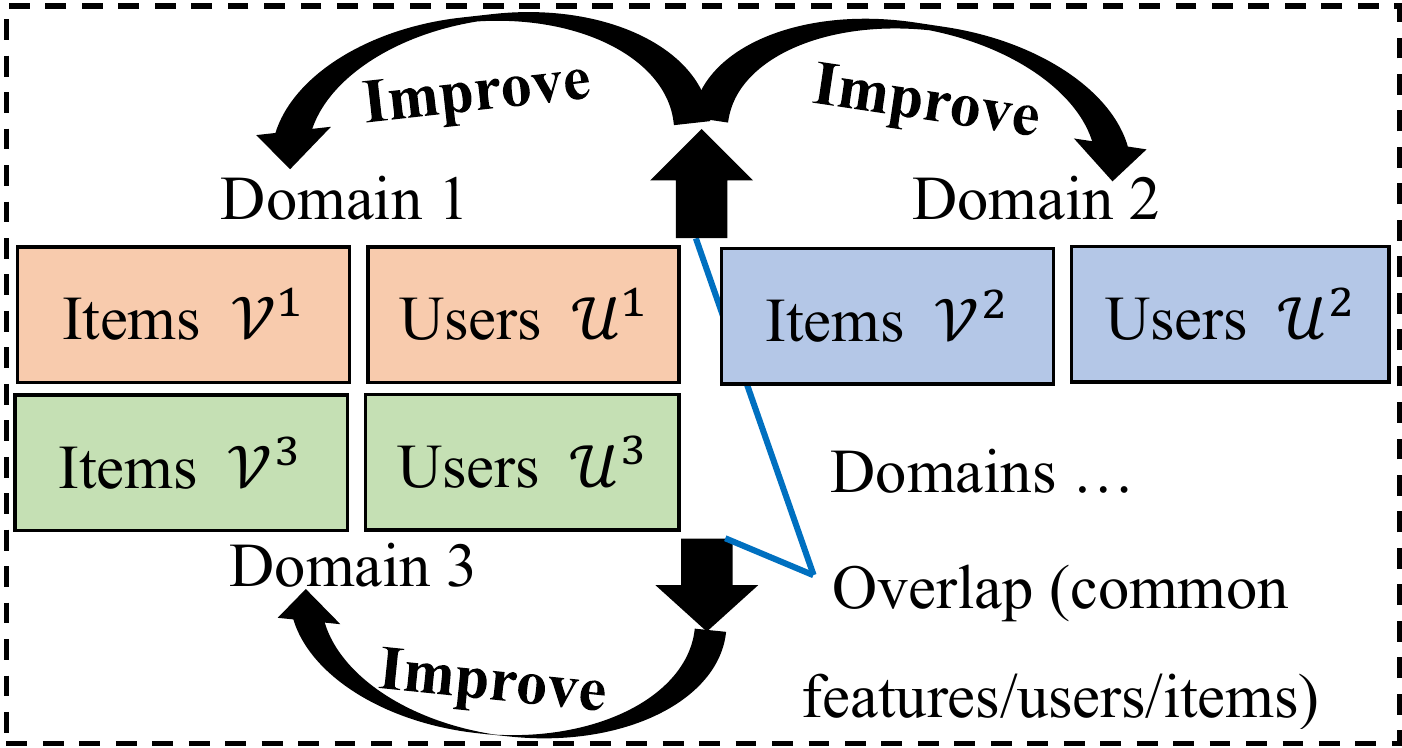}
 		\caption{Multi-target CDR scenario}\label{MTCDR}
 	\end{center}
 \end{figure}
 \begin{figure}[t]
 	\begin{center}
 		\includegraphics[width=0.35\textwidth]{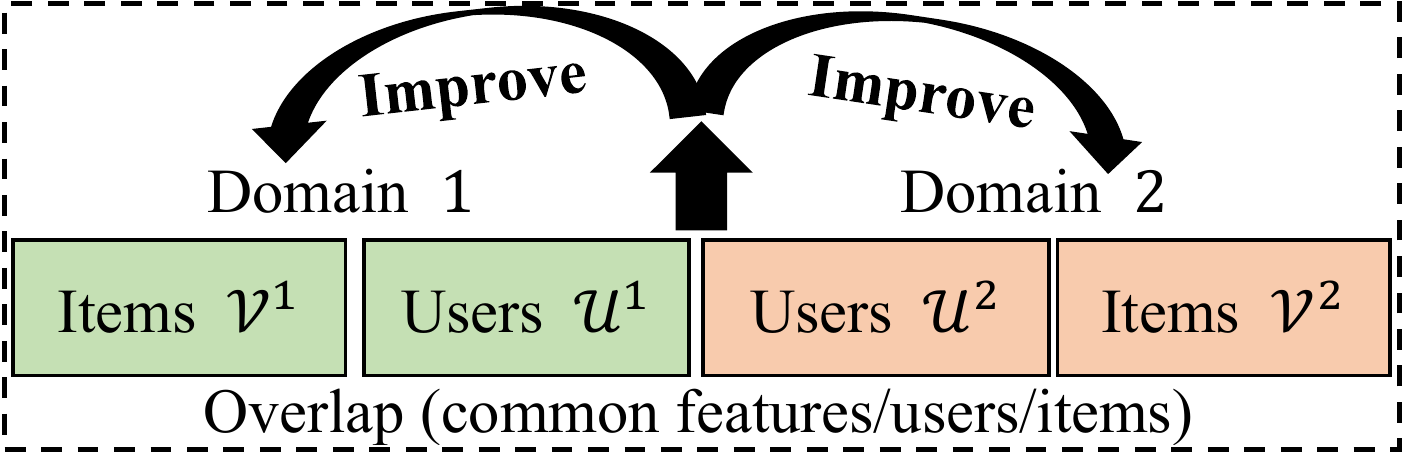}
 		\caption{Dual-target CDR scenario}\label{DTCDR}
 	\end{center}
 \end{figure}
 \begin{figure*}[t]
 	\begin{center}
 		\includegraphics[width=1\textwidth]{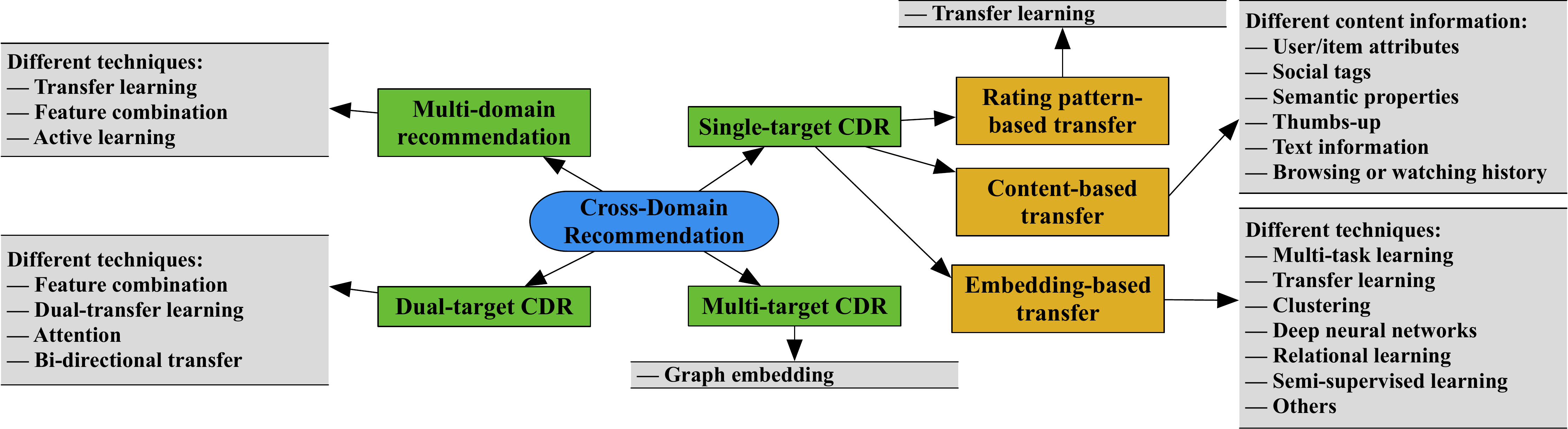}
 		\caption{A categorization of CDR approaches}\label{CDR_Categories}
 	\end{center}
 \end{figure*}
 \begin{table*}[t]
	\caption{The comparison of existing content-based transfer approaches in single-target CDR} \label{STCDR_Existing_CB}
	\begin{center}
	\small{
		\begin{tabular}{c@{   }|@{   }c|@{   }c@{   }|@{   }c@{   }}
			\hline
			\multicolumn{2}{c@{   }|@{   }}{\textbf{Category}}&\textbf{Representative approaches}&\makecell{\textbf{Technology adoption or basic idea}}\\
			\hline
			\multirow{18}{*}{\makecell{\textbf{Content-based} \\ \textbf{transfer}}}&\multirow{2}{*}{\makecell{User/item attributes}}&\cite{berkovsky2007cross}&Multi-source information\\
			\cline{3-4}
			&&\makecell{\cite{leung2007applying}} &User-item \& item-item relationships\\
			\cline{2-4}
			&\multirow{6}{*}{Social tags}
			&\cite{szomszor2008semantic}&Co-occurrence sub-graph\\
			\cline{3-4}
 			&&\cite{abel2011analyzing} &Profile semantic enhancement\\
			\cline{3-4}
			&&\cite{kaminskas2011location} &Tag similarity\\
			\cline{3-4}
			&&\cite{shi2011tags} &Tag similarity\\
			\cline{3-4}
			&&\cite{fernandez2014exploiting} & Rating-tag similarity\\
			\cline{3-4}
			&&\cite{wang2020tag} & Topic modeling\\
			\cline{2-4}
		   &\multirow{4}{*}{Semantic properties}& \cite{fernandez2011generic}&Weighted directed acyclic graph\\
			\cline{3-4}
			&&\cite{kumar2014semantic}&Semantic similarity\\
			\cline{3-4}
			&&\cite{zhang2019cross}&Semantic correlation\\
			\cline{2-4}
			&\multirow{1}{*}{Thumbs-up}& \cite{shapira2013facebook}&\makecell{Preference similarity}\\
			\cline{2-4}
			&\multirow{3}{*}{\makecell{Text information}}&\makecell{ \cite{tang2012cross}}&\makecell{Topic modelling}\\
			\cline{3-4}
			&&\cite{tan2014cross}&\makecell{Topic modelling \& transfer learning}\\
			\cline{3-4}
			&&\cite{sahebi2014content}&\makecell{User similarity}\\
			\cline{2-4}
			&\multirow{2}{*}{\makecell{Browsing or watching history}}&\cite{elkahky2015multi}&\makecell{Multi-view learning}\\
			\cline{3-4}
			&&\cite{kanagawa2019cross}&\makecell{Unsupervised domain adaptation}\\
			\hline
		\end{tabular}
		}
	\end{center}
\end{table*}
\subsection{Multi-Domain Recommendation}
Multi-Domain Recommendation (MDR) is another direction in single-target CDR. It leverages the auxiliary information from multiple domains to recommend a set of items from multiple domains to a certain set of users (single-target) in the multiple domains. We define multiple-domain recommendation as follows.
\begin{myDef} \label{Definition_MDR}
 	\textbf{Multi-Domain Recommendation (MDR)}: Given the multiple domains $1$ to $n$, including user sets $\{\mathcal{U}^1, ... ,\mathcal{U}^n\}$ and item sets $\{\mathcal{V}^1, ... ,\mathcal{V}^n\}$, multi-target CDR is to recommend a set of items $\mathcal{V}^x$ ($\mathcal{V}^x \in \mathcal{V}^1 \cup ... \cup \mathcal{V}^n $) to a certain set of users $\mathcal{U}^x$ ($\mathcal{U}^x \in \mathcal{U}^1 \cup ... \cup \mathcal{U}^n $) and improve the corresponding recommendation accuracy.
\end{myDef}
MDR faces the same challenges as single-target CDR.

 \subsection{Dual-Target CDR}
 Dual-target CDR is a new recommendation scenario in CDR area and it has attracted increasing attention in recent years. We define this recommendation problem as follows.
\begin{myDef} \label{Definition_DTCDR}
 	\textbf{Dual-Target Cross-Domain Recommendation}: Given the two domains $1$ and $2$, including user sets $\mathcal{U}^1$, $\mathcal{U}^2$ and item sets $\mathcal{V}^1$, $\mathcal{V}^2$ respectively, dual-target CDR is to improve the recommendation accuracy in both domains $1$ and $2$ simultaneously by leveraging their observed information.
\end{myDef}

Similar to the problem of single-target CDR, dual-target CDR scenario can be divided into three sub-scenarios according to the notion of domain. This means that dual-target CDR scenarios can also use common features (content-level relevance), common users (user-level relevance), and common items (item-level relevance), to link the two domains and share/transfer knowledge across domains based on these common entities. However, different from single-target CDR, dual-target CDR is to improve the recommendation accuracy in both target domains simultaneously (see Figure \ref{DTCDR}). 
To achieve dual-target CDR, there are two challenges.

\nosection{Designing a feasible dual-target CDR framework (CH4)} 
Unlike conventional single-target CDR, dual-target CDR should enhance the recommendation performance in the two domains, i.e., the source domain and the target domain. Therefore, how to design an effective framework for a dual-target CDR scenario is still very challenging because the auxiliary information from the target domain may negatively affect the performance in the source domain. 

\nosection{Optimizing the embedding of users and items (CH5)} 
In a dual-target CDR scenario, to improve the recommendation accuracy in each domain, the researchers tend to share the common embeddings of common users/items for the two domains or enhance the embedding quality of users/items in each domain by leveraging the auxiliary information from another domain. Therefore, embedding optimization for dual-target CDR scenarios is particularly important. 

 \subsection{Multi-Target CDR}\label{MTCDR_PC}
 Inspired by dual-target CDR, in a multi-target CDR scenario, the researchers aim to improve the recommendation accuracy in multiple domains simultaneously. The core idea of multi-target CDR is to leverage more auxiliary information from more domains to achieve a further improvement of recommendation performance. The problem of multi-target CDR can be defined as follows.
 
 \begin{myDef} \label{Definition_MTCDR}
 	\textbf{Multi-Target Cross-Domain Recommendation}: Given the multiple domains $1$ to $n$, including user sets $\{\mathcal{U}^1, ... ,\mathcal{U}^n\}$ and item sets $\{\mathcal{V}^1, ... ,\mathcal{V}^n\}$, multi-target CDR is to improve the recommendation accuracy in all domains simultaneously by leveraging their observed information. 
\end{myDef}

To achieve multi-target CDR, in addition to the challenges in single-target CDR and dual-target CDR scenarios, there is a new challenge as follows.

\nosection{Avoiding negative transfer (CH6)} 
In a multi-target CDR scenario, the recommendation performance in some domains may decline as more domains, especially sparser domains, join in. This is the \textit{negative transfer} problem that the transferred data/knowledge may negatively affect the recommendation performance in the target domain. In fact, in single-target CDR, MDR, and dual-target CDR scenarios, the researchers may also face the negative transfer problem. However, this problem in multi-target CDR scenarios is more serious because the auxiliary information/knowledge in each domain should be transferred to other domains more than once. Therefore, avoiding negative transfer is an important prerequisite in multi-target CDR scenarios.
	\section{Research Progress}
\begin{table*}[ht]
	\caption{The comparison of existing embedding-based and rating pattern-based transfer approaches in single-target CDR} \label{STCDR_Existing_EB}
	\begin{center}
	\resizebox{0.95\textwidth}{!}{
		\begin{tabular}{@{   }c@{   }|@{   }c|@{   }c@{   }|@{   }c@{   }|@{   }c@{   }}
			\hline
			\multicolumn{2}{@{   }c@{   }|@{   }}{\textbf{Category}}&\textbf{Representative approaches}&\textbf{Training data}&\textbf{Technology adoption or basic idea}\\
			\hline
			\multirow{40}{*}{\makecell{\textbf{Embedding} \\ \textbf{-Based} \\ \textbf{Transfer}}}&\multirow{3}{*}{\makecell{Multi-task\\ learning}}&\cite{singh2008relational}&\makecell{Ratings \& item details} &\makecell{Multiple relations}\\
			\cline{3-5}
			&&\cite{agarwal2011localized}&Multiple contexts &Multi-task learning\\
			\cline{3-5}
			&&\cite{lu2018like}&Ratings &Multi-task learning\\
			\cline{2-5}	
			&\multirow{18}{*}{\makecell{Transfer\\ learning}}
			&\cite{li2009transfer}&Ratings &Transfer learning\\
			\cline{3-5}
			&&\cite{pan2010transfer}&Heterogeneous feedback&Principle coordinates\\
			\cline{3-5}
			&&\cite{zhao2013active}&Ratings&Active transfer learning\\
			\cline{3-5}
			&&\cite{li2014matching}&Ratings &Transfer learning\\
			\cline{3-5}
			&&\cite{wang2016million}&Mails &Mailing list similarity\\
			\cline{3-5}
			&&\cite{zhao2017unified}&Ratings &Active transfer learning\\
			\cline{3-5}
			&&\cite{rafailidis2017collaborative}&Ratings &Transfer learning\\
			\cline{3-5}
			&&\cite{zhang2017cross}&Ratings&Consistent information\\
			\cline{3-5}
			&&\cite{zhang2018cross1}&Ratings &\makecell{Domain adaptation \& \\diffusion kernel completion}\\
			\cline{3-5}
			&& \cite{zhang2018cross2}&Ratings &Feature combination\\
			\cline{3-5}
			&&\cite{hu2018conet}&Ratings &Neural networks\\
			\cline{3-5}
			&&\cite{shang2018demographic}&Ratings &Transfer learning\\
			\cline{3-5}
			&&\cite{he2018robust}&Ratings &Transfer learning\\
			\cline{3-5}
			&&\cite{hu2019transfer}&Ratings \& text & \makecell{Transfer learning \& memory networks}\\
			\cline{3-5}
			&&\cite{manotumruksa2019cross}&Ratings &Transfer learning\\
			\cline{3-5}
			&&\cite{huang2019lscd}&Ratings&Transfer learning\\
			\cline{3-5}
			&&\cite{zhao2020catn}&Ratings&Transfer learning\\
			\cline{2-5}
			&\multirow{4}{*}{\makecell{Clustering}}	&\cite{ren2015improving}&Ratings &Clustering\\
			\cline{3-5}
			&&\cite{rafailidis2016top}&Ratings &User clustering\\
			\cline{3-5}
			&&\cite{farseev2017cross}&Ratings &Clustering\\
			\cline{3-5}
			&&\cite{wang2019solving}&Ratings &Clustering\\
			\cline{2-5}
			&\multirow{6}{*}{\makecell{Deep neural \\networks}}
			&\cite{jaradat2017deep}&Ratings &Textual input relations\\
			\cline{3-5}
			&&\cite{man2017cross}&Ratings &Linear matrix translation \& MLP\\
			\cline{3-5}
			&&\cite{zhu2018deep}&Ratings &Feature combination \& embedding mapping\\
			\cline{3-5}
			&&\cite{he2018general}&Ratings &Bayesian neural networks\\
			\cline{3-5}
			&&\cite{fu2019deeply}&Ratings \& content &Stacked denoising autoencoders\\
			\cline{3-5}
			&&\cite{liu2020exploiting}&Ratings &Aesthetic preferences\\
			\cline{2-5}
			&Relational learning&\cite{sopchoke2018explainable}&Ratings &Relational learning\\
			\cline{2-5}
			&\makecell{Semi-supervised\\ learning}&\cite{kang2019semi}&Ratings &Semi-supervised learning\\
			\cline{2-5}
			&\multirow{5}{*}{Others}
			& \cite{li2011cross}&Ratings &\makecell{Topic model \& interest drift}\\
			\cline{3-5}
			&&\cite{hu2013personalized}&Explicit and implicit feedback &\makecell{Triadic relation (user-item-domain)}\\
			\cline{3-5}
			&&\cite{liu2018transferable}&Ratings &Reinforcement learning\\
			\cline{3-5}
			&&\cite{ma2019pi}&Item sequences &Sequential recommendations\\
			\cline{3-5}
			&&\cite{gao2019cross}&Items' information & Data privacy\\
			\hline
		\end{tabular}
		}
		\resizebox{0.95\textwidth}{!}{
		\begin{tabular}{@{   }c@{   }|@{   }c@{   }|@{   }c@{   }|@{   }c@{   }}
			\hline
			\textbf{Category}&\textbf{Representative approaches}&\textbf{Training data}&\textbf{Technology adoption or basic idea}\\
			\hline
			\multirow{4}{*}{\makecell{\textbf{Rating Pattern-Based Transfer}}}&\cite{li2009can}&\makecell{Ratings} &\makecell{Transfer learning \& cluster-level rating patterns}\\
			\cline{2-4}
			&\cite{gao2013cross}&\makecell{Ratings} &\makecell{Transfer learning \& cluster-level and domain-specific rating patterns}\\
			\cline{2-4}
			&\cite{loni2014cross}&\makecell{Ratings} &\makecell{Interaction patterns}\\
			\cline{2-4}
			&\cite{yuan2019darec}&\makecell{Ratings} &\makecell{Deep learning \& transfer learning}\\
			\cline{2-4}
			\hline
		\end{tabular}
		}
	\end{center}
\end{table*}
 \begin{table*}[t]
	\begin{center}
		\caption{Summary of datasets for CDR}
		\label{Datasets}
		\resizebox{\textwidth}{!}{
		\begin{tabular}{c c c c c}
		\hline
		\textbf{Datasets}& \textbf{Domains} & \textbf{Data types} & \textbf{Scale} & \textbf{Website}\\
		\hline
		Arnetminer \cite{tang2012cross} &\makecell{Research domains (user-level\\ relevance --- attribute-level)}&\makecell{Paper \& author \& \\conference name ...}&1 million & \url{https://www.aminer.org/collaboration}\\
		\hline
		\makecell{MovieLens + Netflix\\ \cite{zhu2018deep}} & \makecell{Movie (item-level relevance)}& Rating \& tag & \makecell{25 million \\\& 100 million} & \makecell{\url{https://grouplens.org/datasets/movielens/}\\ \url{https://www.kaggle.com/netflix-inc/netflix-prize-data}}\\
		 \hline
		Amazon \cite{fu2019deeply}  &\makecell{Book \& music \& movie ... \\(user-level relevance --- type-level)} &\makecell{Rating \& review \\\& side information} & 100 million+& \url{http://jmcauley.ucsd.edu/data/amazon/} \\
		\hline
		Douban \cite{zhu2020graphical}  &\makecell{Book \& music \& movie \\(user-level relevance --- type-level)} &\makecell{Rating \& review \\\& side information} & 1 million+& \url{https://github.com/FengZhu-Joey/GA-DTCDR/tree/main/Data} \\
		\hline
		\end{tabular}
	   }
	\end{center}
\end{table*}
To correspond with the recommendation scenarios and challenges mentioned in Section \ref{Section_PC}, in this section, we summarize the existing CDR approaches according to their target scenarios, target challenges, data categories, and technical perspectives. We also summarize the popular datasets in CDRs.
 
\subsection{Single-Target CDR}\label{RP_STCDR}
Most of existing single-target CDR approaches tend to leverage useful information from the source domain to the target domain. According to transfer strategies, these single-target CDR approaches are divided in three categories: content-based transfer, embedding-based transfer, and rating pattern-based transfer.

\nosection{Content-Based Transfer} 
To target \textbf{CH1}, the existing content-based transfer approaches first create links based on the common contents, e.g., user/item attributes, social tags, semantic properties, thumbs-up, text information, metadata, and browsing or watching history. 
Then, they transfer user/item data or knowledge across domains. We clearly compare the difference of these approaches in Table \ref{STCDR_Existing_CB}.

\nosection{Embedding-Based Transfer} 
To target \textbf{CH2} and \textbf{CH3}, the existing embedding-based transfer approaches employ some classical machine learning models, e.g., multi-task learning, transfer learning, clustering, deep neural networks, relational learning, and semi-supervised learning, to map or share embeddings, e.g., user/item latent factors, learned by CF-based models across domains. In addition to these frequently-used learning techniques, other embedding-based transfer approaches tend to employ different techniques or ideas, e.g., Bayesian latent factor models \& interest drift, triadic relation (user-item-domain), reinforcement learning, sequential recommendations, and data privacy. We clearly compare the differences among these approaches in Table \ref{STCDR_Existing_EB}.

\nosection{Rating Pattern-Based Transfer} 
To target \textbf{CH2}, the existing rating pattern-based transfer approaches tend to first learn an independent rating pattern of users from the source domain and then transfer the rating pattern for the target domain to improve the corresponding recommendation accuracy. 
The representative work of this type of approach includes 
\cite{gao2013cross,he2018robust,yuan2019darec}. We list the difference of these approaches in Table \ref{STCDR_Existing_EB}.

 \subsection{Multi-Domain Recommendation}\label{RP_MDR}
Multi-Domain Recommendation (MDR) is another direction in single-target CDR, but it achieves a different goal: it makes recommendations for different domains. Some of these multi-domain approaches can be applied in CDR scenarios, but they tend to make recommendations either for specific or common users who are selected from domains, or for the users in the target domain only.

 MDR also faces the conventional challenges, e.g., \textbf{CH1} and \textbf{CH2}, in STCDR. To address these challenges, in \cite{zhang2012multi}, Zhang et al. proposed a multi-domain collaborative filtering (MCF) framework for solving the data sparsity problem in multiple domains. After this, the MDR models proposed in \cite{cao2010transfer,moreno2012talmud,pan2013transfer,zhang2016multi} employ different techniques, i.e., feature combination, transfer learning, and active learning to transfer the knowledge of similar/common users among multiple domains.

 \subsection{Dual-Target CDR}
 Dual-target CDR is still a novel but very attractive concept for improving the recommendation accuracies in both domains simultaneously. Therefore, existing solutions are limited but the researchers are paying more and more attention to this direction. To target \textbf{CH4} and \textbf{CH5}, the existing dual-target CDR approaches mainly focus on either applying fixed or flexible combination strategies \cite{zhu2019dtcdr,zhu2020graphical,liu2020cross}, or simply changing the existing single-target transfer learning to become dual-transfer learning \cite{li2019ddtcdr}.

In \cite{zhu2019dtcdr}, Zhu et al. first proposed the DTCDR, a dual-target CDR framework that uses multi-source information such as ratings, reviews, user profiles, item details, and tags to generate more representative embeddings of users and items. Then, based on multi-task learning, the DTCDR framework uses three different combination strategies to combine and share the embedding of common users across domains. Similarly, in \cite{liu2020cross}, Liu et al. also use a fixed combination strategy, i.e., hyper-parameters and data sparsity degrees of common users, to combine the embedding of common users.

Additionally, a new dual-target CDR model (DDTCDR) was proposed in \cite{li2019ddtcdr}, which considers the bi-directional latent relations between users and items and applies a latent orthogonal mapping to extract user preferences. Based on the orthogonal mapping, DDTCDR can transfer users' embeddings in a bidirectional way (i.e., Source $\rightarrow$ Target and Target $\rightarrow$ Source). Recently, Zhu et al. proposed another dual-target CDR framework in \cite{zhu2020graphical}, which employs graph embedding to generate more informative embeddings of users and items, and employs element-wise attention to combine the embeddings of common users/items across domains.
\subsection{Multi-Target CDR}
Although multi-target CDR is inspired by dual-target CDR and multi-domain recommendation, it aims to achieve a bigger goal, i.e., providing a complete solution for data sparsity. In principle, if the multi-target CDR models can find enough related domains and utilize the auxiliary information from these multiple domains well, the long-standing data sparsity problem in recommender systems can be greatly alleviated and even solved. However, as introduced in Section \ref{MTCDR_PC}, apart from the challenges in single-target CDR and dual-target CDR scenarios, a new challenge, i.e., negative transfer (\textbf{CH6}), is inevitable in real multi-target CDR scenarios.
 
Multi-target CDR is a challenging recommendation scenario, and thus, by now, there are few solutions \cite{cui2020herograph,krishnan2020transfer} on achieving this goal. In \cite{cui2020herograph}, the authors use a shared heterogeneous graph to generate more informative embeddings of users and items among multiple domains. Also, the MDCDR approach proposed in \cite{krishnan2020transfer} leverages the auxiliary information from a source domain to improve the recommendation accuracy of multiple domains. However, these approaches do not consider the negative transfer problem. Therefore, multi-target CDR is still a challenging task in CDR.

\subsection{Summary of Datasets}
In this section, we summarize several popularly used datasets for CDR tasks in Table \ref{Datasets}. This will guide the researchers to obtain these CDR datasets conveniently. Anyone who wishes to use these datasets can refer to the corresponding citations and websites for more details.

	\section{Research Prospects}
Although many efforts have been put to tackle the challenges of CDR, there remain some promising prospects, and we summarize three of them as follows. 

\nosection{Heterogeneous CDR}
Most existing CDR approaches assume information across domains is homogeneous, which is not consistent with reality. 
For example, some researchers assume both domains have rating and content information \cite{winoto2008if}, while other studies assume the existence of rating data across both domains \cite{zhao2017unified}. 
However, in practice, different domains are rich in different kinds of information. 
For instance, an e-commerce domain (e.g., Amazon) is rich in user-item interaction data while a social domain (e.g., Facebook) has plenty of user-user social data. 
Under such situations, new techniques should be proposed to identify the `bridges' across domains so as to transfer information and improve the performance of CDR. 
How to leverage these heterogeneous data across domains, to further improve the recommendation performance, becomes the first promising prospect in CDR. 

\nosection{Sequential CDR}
Sequential recommendation has gained much attention since it can suggest items to users by modeling the sequential dependencies over the user-item interactions \cite{wang2019sequential}. 
Naturally, CDR also faces the problem of sequentially modeling of users and items, the same as conventional recommender systems. 
Prior work on sequential recommendation mainly focuses on learning the high-order, long-term, and noisy user-item interactions in sequence. 
It becomes more challenging for sequential CDR since one not only needs to model sequential user-item interactions, but also transfer information across domains \cite{ma2019pi}. 
Therefore, sequential CDR becomes the second promising research prospect. 
  
\nosection{Privacy-Preserving CDR}
Most existing approaches in CDR assume that data across domains are available in plaintext, which ignores the data isolation problem in practice. 
Apparently, most recommender systems are built using users' sensitive data, e.g., check-in data, user profile, and browse history. And in CDR, these data are usually held by different domains, e.g., Amazon and eBay. 
In some cases, these data across domain cannot share with other directly since they contain sensitive information. 
Thus, it is urgent to build CDR meanwhile protect data privacy \cite{gao2019cross}. 
A recent study on privacy-preserving CDR can only handle the simple social matrix factorization model \cite{chen2019secure}, and there is a long way to go for complex privacy-preserving CDR. 
Therefore, privacy-preserving CDR is the third pressing and promising research prospect.

	\section{Conclusion}
Cross-domain recommendations (CDRs) have attracted increasing research attention, with the development of deep neural network and graph learning techniques. 
This paper conducted a comprehensive survey on the following four scopes, i.e., 
single-target CDR, multi-domain recommendation, dual-target CDR, and multi-target CDR. 
We first have presented the definitions and challenges of these scopes, then proposed a full-view categorization and new taxonomies on these scopes, and finally listed several promising prospects in CDR. 
This survey summarizes current representative research efforts and trends and we expect it can facilitate future research in the community.

\newpage
\bibliographystyle{named}
\bibliography{ijcai21}	
\end{document}